\documentclass[]{spie}  

\usepackage{amsmath,amsfonts,amssymb}
\usepackage{graphicx}
\usepackage[colorlinks=true, allcolors=blue]{hyperref}
\usepackage{rotating}
\usepackage{subcaption}
\usepackage[export]{adjustbox}

\title{On-sky commissioning and technical performance of Proto-NUX at Pic du Midi}

\author[a]{Rasjied Sloot}
\author[a]{Rudy Wijnands}
\author[b]{Steven Bloemen}
\author[a]{Hans Ellermeijer}
\author[a]{Alexander Hoogerbrug}
\author[c]{Rik ter Horst}
\affil[a]{Anton Pannekoek Institute for Astronomy, University of Amsterdam, Science Park 904, 1098 XH, Amsterdam, The Netherlands}
\affil[b]{Department of Astrophysics/IMAPP, Radboud University, PO Box 9010, 6500 GL Nijmegen, The Netherlands}
\affil[c]{NOVA Optical IR Instrumentation Group at ASTRON, Oude Hoogeveensedijk 4,  7991 PD Dwingeloo, The Netherlands}

\authorinfo{Further author information: (Send correspondence to Rasjied Sloot)\\Rasjied Sloot: E-mail: m.r.sloot@uva.nl}

\begin{document} 
\maketitle

\begin{abstract}
The Near-Ultraviolet eXplorer (NUX) is a proposed ground-based observing facility designed to expand near-ultraviolet (NUV) astronomy by enabling observations in the 300--350 nm range from Earth. The full system is intended to consist of four modified 36 cm Celestron RASA telescopes with custom NUV-optimized optics and UV-sensitive CMOS detectors, providing a total field of view of about 67 square degrees for high-cadence searches for fast, hot transients. To validate this concept, a single-telescope prototype, Proto-NUX, was constructed. Its optical design and development are described in previous work. Initial sea-level tests in the Netherlands showed that the instrument could be operated successfully and that the main technical requirements were met. In March 2026, Proto-NUX was deployed at the Pic du Midi Observatory in the French Pyrenees, at an altitude of 2877 m, for its first high-altitude on-sky commissioning campaign. The campaign tested the full system under realistic observing conditions, including deployment, remote operation, calibration procedures, detector behaviour, image quality, tracking, ghosting, and simultaneous optical monitoring. The observations demonstrate that Proto-NUX is a viable platform for ground-based NUV observations, but also show that the system was not yet operating at its expected optical limit. In particular, the delivered NUV point-spread function was broader than expected, with likely contributions from residual collimation error, defocus, chromatic effects, and internal reflections. The detector dark current was also found to be a significant contributor to the noise budget. We discuss these limitations and identify the main improvements needed for future Proto-NUX campaigns and for scaling the concept to the full NUX facility.
\end{abstract}


\section{Introduction} \label{sec:intro}
 
Ground-based astronomy is often described in terms of atmospheric windows: wavelength ranges where the atmosphere is transparent enough for viable observations from the Earth's surface. The 300--350 nm near-ultraviolet (NUV) band lies close to the short-wavelength edge of this ground-accessible window. In this regime, the practical sensitivity is set by the combined atmospheric transmission, detector response, optical throughput, and sky background, each of which changes rapidly with wavelength. Ozone absorption becomes especially important toward the blue end of the band, while Rayleigh scattering and aerosol extinction also contribute to the total atmospheric extinction (see, e.g., [\citenum{Bucholtz1995,Orphal2016,2011A&A...527A..91P}]). The 300--350 nm band is therefore not simply a bluer extension of optical $u$-band observations. It is a separate observing regime, whose useful wavelength range and practical sensitivity have to be measured on sky (see discussion in [\citenum{2024SPIE13094E..30W}]).

The scientific motivation for observations in this band is broad. NUV emission traces hot astrophysical sources. Many transients, in particular, are expected to emit strongly at NUV wavelengths during the first hours to days after explosion or outburst. Examples include the shock-breakout and shock-cooling phases of supernovae, early gamma-ray burst afterglows, blue kilonova emission from neutron star - neutron star mergers, tidal disruption events, and other fast blue transients (see, e.g.,  [\citenum{Waxman2017,Nakar2010,Oates2009,Nicholl2017,Sagiv2014}]). A wide-field, high-cadence survey in the NUV can probe the physical characteristics of such objects (e.g.,
temperature, radius) in ways that are difficult to infer from optical data alone. Space-based UV telescopes avoid the atmospheric cutoff and have provided much of the current view of the UV sky, for example with GALEX\cite{Martin2005} and the UV-optical telescope\cite{Roming2005} aboard the Neil Gehrels Swift observatory. However, wide-field UV time-domain facilities remain rare, although ULTRASAT\cite{Sagiv2014,Shvartzvald2024} is expected to open this parameter space from orbit in the 230–290 nm wavelength range after its planned launch in late 2027 or early 2028. A ground-based instrument operating in the accessible 300--350 nm window therefore offers a complementary approach, provided that atmospheric transmission and sky background are accurately characterised.

The Near-Ultraviolet eXplorer (NUX)\cite{2024SPIE13094E..30W} was proposed to make use of this ground-accessible NUV window. The full instrument is planned as an array of four modified 36 cm Celestron RASA telescopes with NUV-optimized optics and UV-sensitive CMOS detectors. Together, these telescopes will cover $\sim$67 square degrees in the 300--350 nm range. The aim is to reach an AB magnitude of 20 in a 150 s exposure under dark conditions, which would allow high-cadence searches for the fast, hot transients discussed above. Whether this can be achieved from the ground depends not only on the optical design and the detector sensitivity, but also on the atmospheric transmission, sky background, photometric stability, and achievable image quality.

Proto-NUX\cite{2026PASP..138e5004S} was developed as a single-telescope pathfinder for the NUX concept. It uses one of the telescopes planned for the full facility, but with the optical train adapted for NUV observations. The main changes include a NUV-transparent Schmidt corrector\footnote{For the construction of this corrector, we refer to [\citenum{2024SPIE13094E..30W}].}, a custom NUV lens assembly, a recoated primary mirror, and a UV-sensitive CMOS detector. The instrument can be used with three filter configurations: the 300--350 nm full-NUX band, the 300--325 nm short-NUX band, and the 325--350 nm long-NUX band. These sub-bands were chosen to separate the shorter-wavelength part of the band, where ozone absorption is important, from the longer-wavelength part, where Rayleigh scattering is expected to dominate over ozone absorption. During the campaign, a complementary optical telescope was mounted in parallel to provide simultaneous optical coverage and to support calibration and performance checks.

This paper presents the first high-altitude on-sky commissioning
campaign of Proto-NUX, carried out at the Pic du Midi Observatory in March 2026. The campaign was intended to move beyond laboratory tests and sea-level observations, and to test the full system under realistic observing conditions. We focus on the technical and operational performance of the instrument: deployment at the site, observing workflow, detector behaviour, focusing and collimation, image quality, tracking, ghosting, calibration strategy, and the practical lessons for future Proto-NUX campaigns and for the full NUX facility. A detailed analysis of the atmospheric extinction, its temporal variability, and the NUV sky background will be presented separately\cite{protonuxaapaper}. Here, we use the commissioning data to assess whether Proto-NUX is a viable platform for ground-based NUV observations, and to identify the changes needed before future observing campaigns.

\section{Proto-NUX campaign} \label{sec:protonuxcampaign} 

The first stage of the Proto-NUX program consisted of procurement, modification, assembly, and initial commissioning of the instrument, as described by [\citenum{2026PASP..138e5004S}], where the Amsterdam pre-campaign tests were briefly mentioned as well. Here we describe those tests in more detail and present the full on-site testing carried out in March 2026 at the Pic du Midi Observatory in the French Pyrenees. Describing this campaign is the main purpose of this paper. The next stage will consist of further long-term testing and is outlined in Section \ref{sec:future}.

\subsection{Pre-campaign testing in Amsterdam} \label{sec:amsterdam}

Before the observing campaign at Pic du Midi, we had two opportunities for test observations at the Anton Pannekoek Observatory\cite{APO_paper} at the University of Amsterdam (specifically on 24 January and 4 March, 2026). No further test observations could be carried out before the Pic du Midi campaign because the telescope assembly was completed only in mid-January 2026 and these were the only two clear-night opportunities before the already scheduled Pic du Midi campaign. 

\begin{figure}
    \centering
    \includegraphics[width=7.4cm,angle=-90,valign=c]{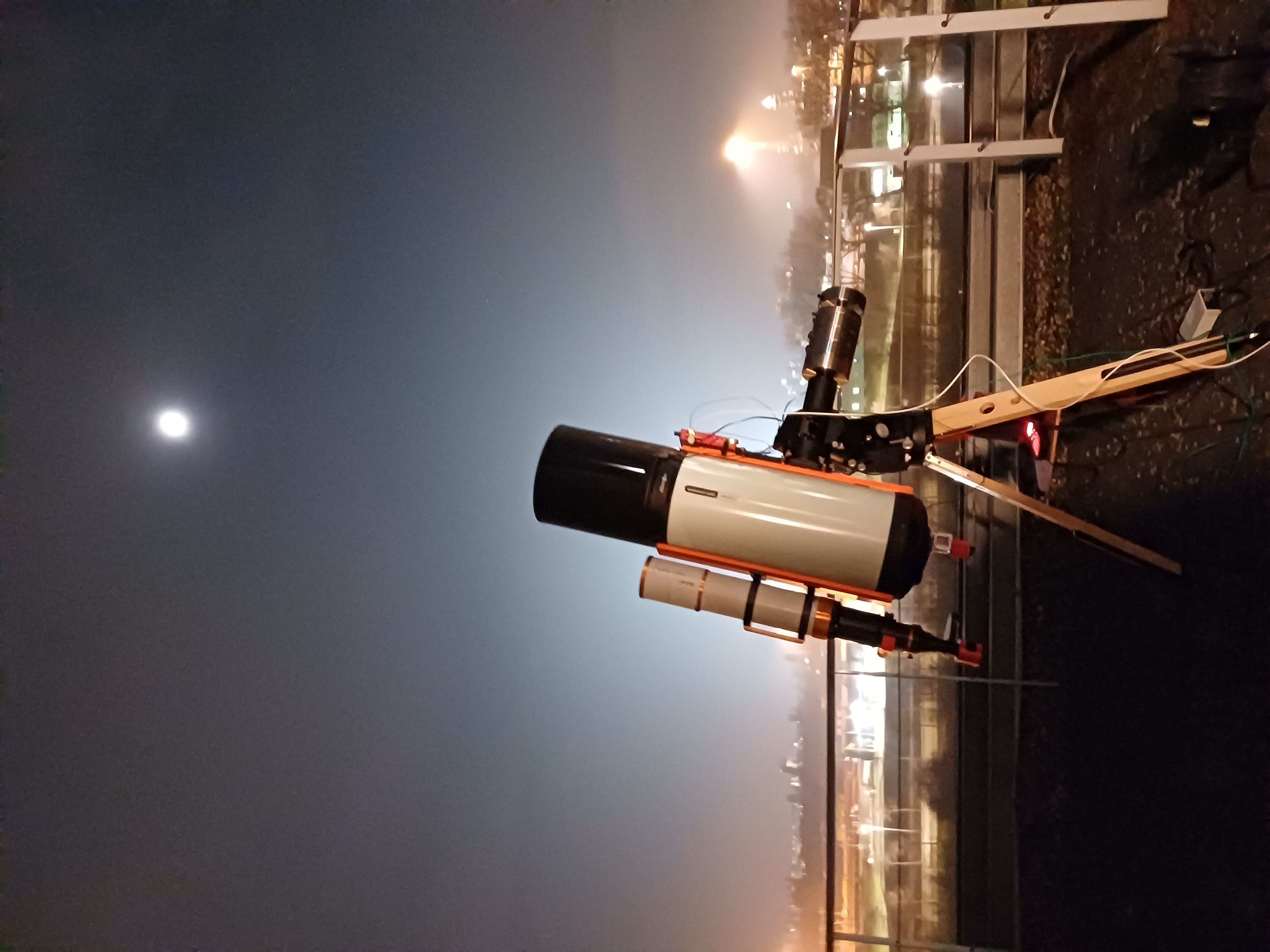}
    \hspace{0.1cm}
    \includegraphics[height=7.4cm,valign=c]{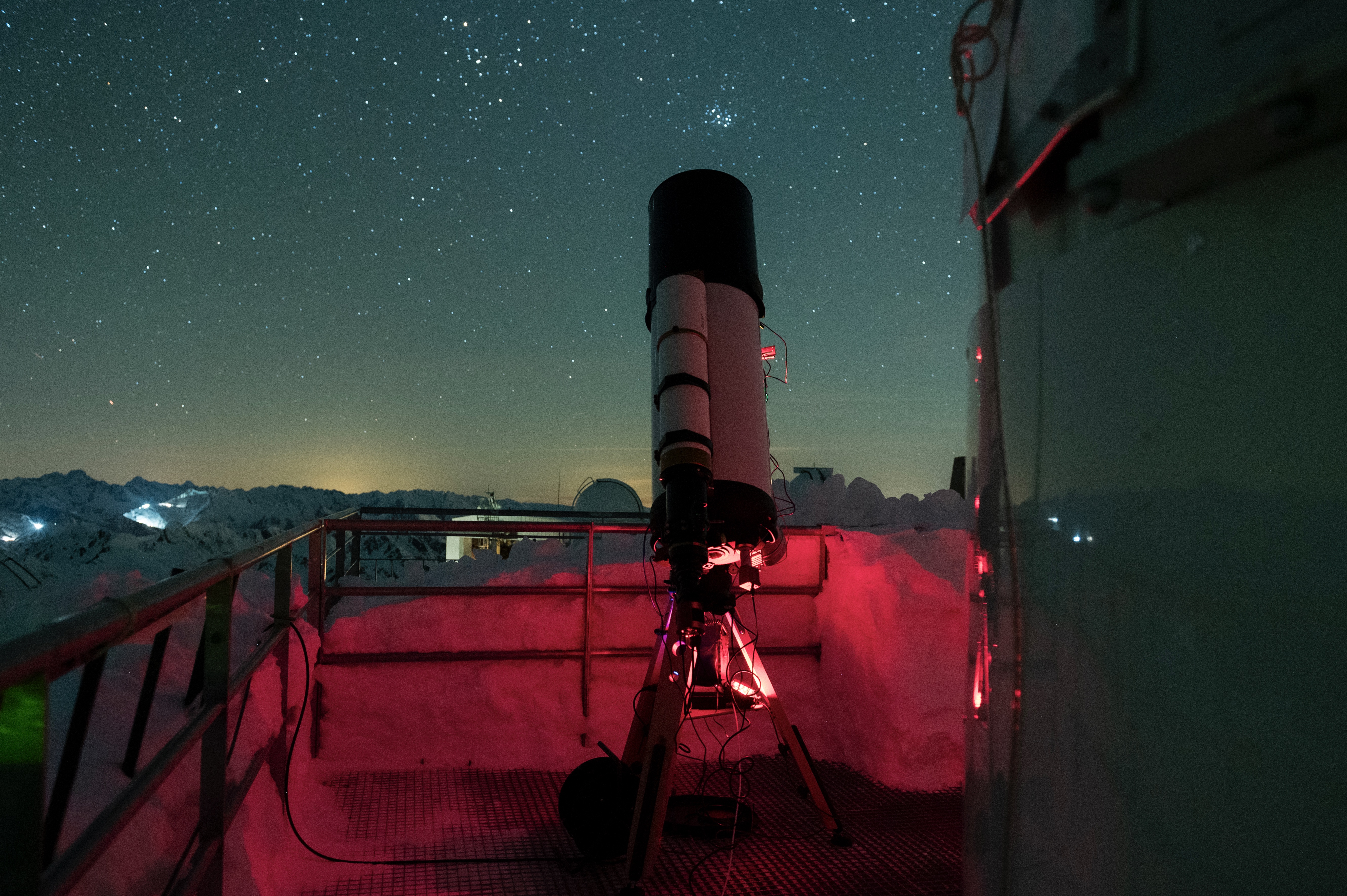}
    \vspace{0.25cm}
    \caption{{\it Left:} Proto-NUX during the first test observations on 24 January 2026, installed on the roof of the Anton Pannekoek Observatory at the University of Amsterdam. The hazy observing conditions are evident from the strong scattering around the Moon near the top of the image. {\it Right:} Proto-NUX during the first observing night on 12 March 2026 on the designated platform on Pic du Midi (photo credit Zara Nor).}
    \label{fig:aponight}
\end{figure}

The first night was very hazy (see Fig.~\ref{fig:aponight}, left), with only a handful of stars visible with the naked eye. The main objectives were therefore only technical: to test the basic operations of the setup, including telescope installation, mount balancing and modeling, assembling of the filter configurations, remote operation of all equipment, and the acquisition of test images of on-sky targets using the three Proto-NUX filters. During the first test night, the initial image quality revealed clear optical alignment issues, which turned out to be caused by severe sensor tilt: Within the same field, both in-focus and out-of-focus stellar images were present. This was traced to a mechanical issue in the connection between the filter slider and the lens assembly.  During the second test night, a few days before departure to Pic du Midi, the main objectives were to verify the correction of the sensor-tilt problem and to finalize the packing list for the observing campaign.  A photo of the setup during this second test night is shown in [\citenum{2026PASP..138e5004S}]. An initial inspection indicated that the focus and collimation were at an acceptable level for the start of the Pic du Midi campaign.

\subsection{Site deployment and logistics} \label{sec:logistics}

The Pic du Midi Observatory\footnote{https://picdumidi.com/en} is located in the French Pyrenees at an altitude of 2877 meter.  Its high elevation, dark skies, and generally good observing conditions make it a suitable site for testing Proto-NUX under reduced atmospheric column density compared with the sea-level observations taken in Amsterdam (Section~\ref{sec:amsterdam}).

Accommodation for visiting astronomers was available in the Dauz\`ere-Soler building and was arranged through the University of Toulouse. For our campaign, accommodation and access the observing platform on the sixth floor of the building (see Fig.~\ref{fig:picdumidioverview}) were booked for 9--20 March 2026. During the stay, a room on the fifth floor was made available for unpacking, temporary storage, and preparation of the equipment. This nearby indoor space was essential for assembling and checking the instrument before moving it to the observing platform. 

The Pic du Midi observing campaign required the full Proto-NUX setup to be transported from the Netherlands to La Mongie, France, by car and subsequently transferred to the observatory by cable car, or t\'el\'eph\'erique. All equipment was packed in transportable cases, with the Proto-NUX optical tube assembly carried in a dedicated wheeled flight case. The cable car reaches the summit from La Mongie in approximately 15 minutes, including an intermediate stop. It normally transports visitors, but can also be used to carry scientific equipment, subject to operational scheduling and weight constraints.  Although the main equipment cases had wheels, handling remained challenging under winter conditions. In particular, heavy snow at the summit made transport from the cable-car arrival point to the observing building difficult.

Since the Proto-NUX setup (Section~\ref{sec:setup}) was designed for remote operation, the installation requirements at the site were limited to a stable platform, electricity, and a wired Ethernet. Batteries were available as backup in case of power outages, but none occurred during our campaign. Electricity was already present on the platform and an Ethernet cable was routed through an existing wall opening by the observatory staff. Upon arrival, snow was present on the platform and was removed by the observatory staff. Following additional snowfall during the observing campaign, the platform was cleared again during daytime operations. During clear-weather intervals, the full setup remained installed on the observing platform during daytime and was protected by a weatherproof tarp. When snowfall was expected, the setup was disassembled and moved indoors. It was reinstalled when conditions allowed, i.e. when skies were clear and no further snowfall was forecast.

The return transport also required planning around cable-car availability. Because Pic du Midi is a popular destination for tourists and school visits, the cable car was frequently occupied during daytime operations, with priority given to scheduled visitor transport; therefore, return transport of the equipment had to be coordinated with the observatory staff in advance.

\subsection{Instrument configuration during the campaign} \label{sec:setup}

\begin{figure}
    \centering
    \includegraphics[width=1\linewidth]{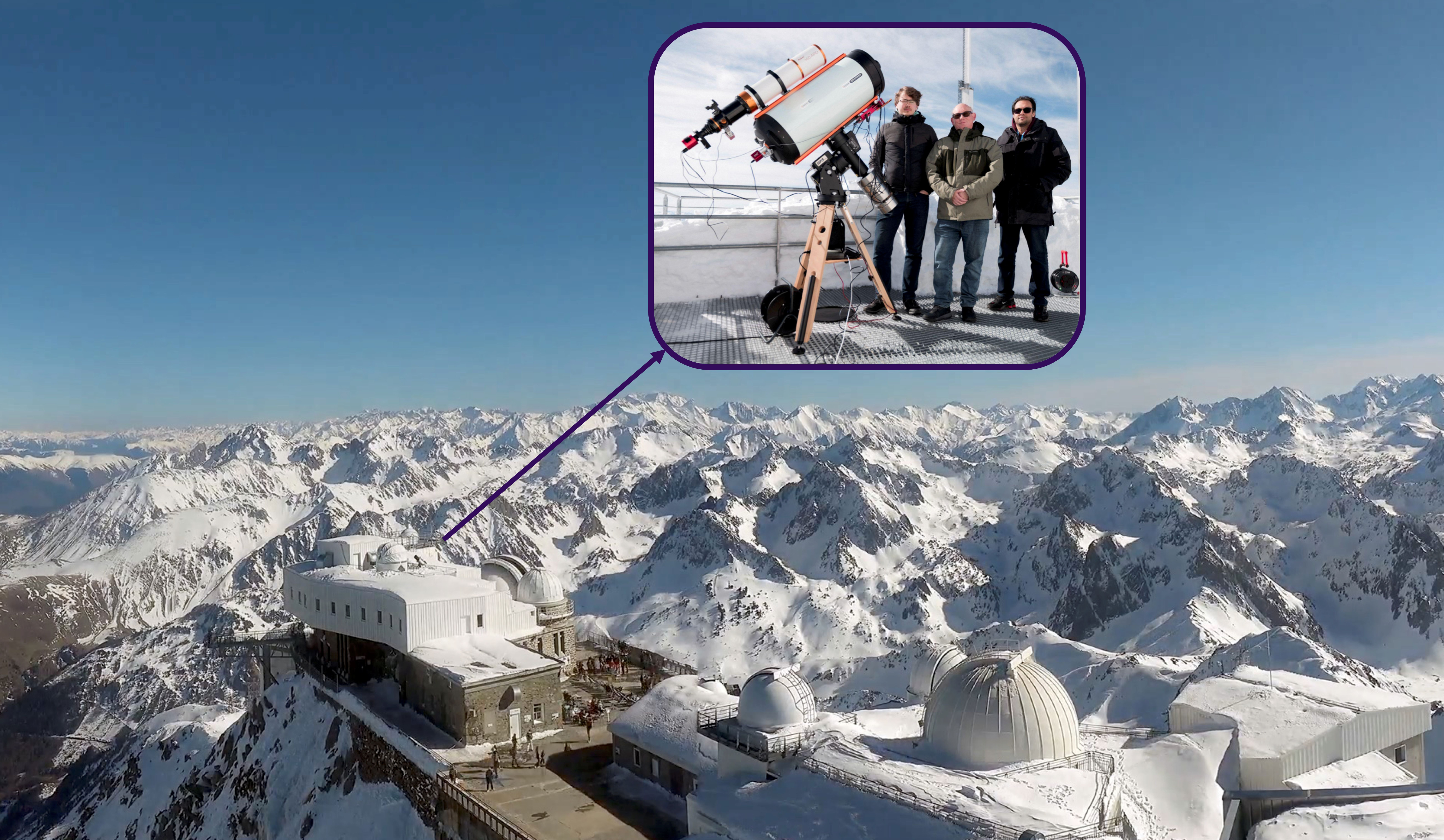}
    \vspace{0.25cm}
    \caption{An overview of the Pic du Midi Observatory on 18 March 2026, with the observing platform for Proto-NUX visible in the background. The inset shows Proto-NUX on the platform during the campaign, together with the three observers: Alexander Hoogerbrug, Rudy Wijnands, and Rasjied Sloot (from left to right). Overview photo obtained from https://picdumidi.com/en/live/ and inset photo credit by Zara Nor.}
    \label{fig:picdumidioverview}
\end{figure}

The observing setup deployed at Pic du Midi consisted of Proto-NUX, a Complementary Optical eXplorer (COX), a 10Micron equatorial mount, local control hardware, and auxiliary sky-monitoring equipment. Proto-NUX is a modified 36 cm aperture telescope based on the Celestron RASA design. The details of the optical redesign, filter concept, and expected system throughput of Proto-NUX  have been described in [\citenum{2026PASP..138e5004S}]. The secondary telescope, COX, was a 14 cm aperture apochromatic triplet refractor (ASKAR 140) mounted in piggyback configuration on Proto-NUX (see Fig.~\ref{fig:picdumidioverview}). It was attached to the Losmandy-style dovetail rail on top of the Proto-NUX optical tube and was aimed to observe the same field simultaneously. COX was equipped with a ZWO electronic filter wheel containing seven 1.25-inch threaded SDSS filters ($u$, $g$, $r$, $i$, $z$, $y$) and a luminance filter. Because the piggyback mounting arrangement did not provide fine adjustment of the relative alignment between the two optical axes, a residual pointing offset of 36 arcminutes remained between the two telescopes. The overlapping field of view was therefore about one quarter of the Proto-NUX field, corresponding to approximately 0.2 square degrees.

The full telescope assembly was mounted on a 10Micron GM2000 HPS II Combi equatorial mount. The mount uses high-precision absolute encoders and a pointing model to provide accurate tracking without an autoguiding feedback system. The total payload of the deployed setup was approximately 50 kg. Owing to the piggyback configuration of the secondary telescope, the full counterweight bar had to be used, carrying five 12 kg counterweights and one 6 kg counterweight.

Proto-NUX was equipped with a QHY2020 camera using the back-side illuminated GSENSE2020BSI CMOS sensor with 2048 $\times$ 2048 pixels. The camera was operated in dual-readout mode, in which each exposure is read out through both a low-gain and a high-gain ADC. The resulting raw image has dimensions of 2048 $\times$ 4096 pixels, with the low-gain channel on one side and the high-gain channel on the other. During the observing campaign, the camera was cooled to $-10^{\circ}$C. This temperature was chosen to provide stable detector performance while still allowing matching dark frames to be acquired indoors during daytime calibrations using the thermoelectric cooler.

Proto-NUX used four manually exchangeable 50 $\times$ 50 mm filter-slider configurations. These included the full-NUX, short-NUX, and long-NUX configurations\cite{2026PASP..138e5004S}, together with an additional $u$-band configuration used for testing and comparison with $u$-band observations with COX. Because the filters were mounted in separate sliders, filter changes required manually moving the telescope to an accessible position before exchanging the slider. The use of absolute encoders on the mount allowed the telescope to be moved manually for filter changes without losing the pointing model, enabling observations to resume quickly after each exchange.

Focusing of Proto-NUX was performed using custom-made tri-Bahtinov masks. Two versions were produced: one fitted over the dew shield and one fitted directly to Proto-NUX without the dew shield. The latter included a central aperture to accommodate the front-mounted camera. These masks were used during setup and focus checks on bright stars.

The sky background was monitored with a Unihedron Sky Quality Meter LE mounted on a fence at the observing platform and pointed toward zenith. This device recorded the broadband sky brightness in mag arcsec$^{-2}$ throughout the night, providing an independent monitor of sky-background variations during the observations.

A PrimaLuceLab Eagle 5 XTM was used as the control PC for the observing system. It controlled the cameras, mount interface, and 12 V power distribution for the cameras and auxiliary hardware. The Eagle was mounted on the lower Losmandy-style rail of the Proto-NUX tube. A local network switch connected the Eagle, the mount, the sky-quality meter, and a laptop used for remote-desktop access over a local LAN. The observing setup was operated without internet connectivity; for a permanent installation, a similar local-control configuration with reliable remote access would be preferred.

\subsection{Observing schedule}

The Pic du Midi observing campaign was carried out in March 2026. We arrived at the observatory in the afternoon of March 11. Weather conditions on the first night were unsuitable for observations due to dense fog. The time was therefore used to unpack the equipment and prepare the instruments for deployment.

On March 12, the weather cleared during the afternoon, allowing the mount and telescopes to be installed on the observing platform. The first observations were obtained that night (Fig.~\ref{fig:aponight}, right). During the first half of the night, however, strong winds led to large and non-uniform stellar profiles in the images. No observations were possible on March 13 and 14 because of poor weather conditions, including heavy snowfall and  strong winds, and the full setup was moved indoors into the control room during this period. Clear conditions returned on March 15 and continued through March 18, providing four consecutive observing nights. During this interval, the setup could remain installed on the observing platform between nights under the weatherproof tarp described above (Section ~\ref{sec:logistics}).

On March 19, heavy snowfall was again expected during the afternoon. The setup was therefore moved indoors. Given the uncertain forecast for the final possible observing night and the fact that sufficient data had already been obtained to meet the main objectives of the campaign, we decided to pack all equipment and leave the observatory on the afternoon of March 19. A summary log of the observations is given in Table~\ref{tab:Observations}.

\subsection{Operational workflow} \label{sec:workflow}

A typical observing night started with powering the system, establishing the local network connection to the control PC, and cooling the cameras to a $-10^{\circ}$C operating temperature. The cooling sequence was started before evening twilight to allow the detector temperature to stabilize before calibration frames and science observations were acquired.

Twilight sky flats were obtained near zenith before the start of the science program. For Proto-NUX, flats were acquired in the manual filter-slider configurations in the order short-NUX, long-NUX, full-NUX, and U. This order was chosen to obtain sufficient signal in the lowest-throughput configurations while the twilight sky was still bright, before moving to the higher-throughput filters at lower sky brightness. For COX, flats were acquired using the electronic filter wheel in the order $y$, $u$, $z$, $i$, $r$, $g$, $L$. Typical exposure times were adjusted during the sequence to maintain count levels within the linear detector regime.

For COX, filter changes during both flat-field acquisition and science observations were performed remotely from the control room through the local network connection to the Eagle control PC and its electronic filter wheel. Complete multi-filter sequences could therefore be run without manual intervention at the telescope. In contrast, changing the Proto-NUX filters had to be performed manually at the telescope, both during the flat-field sequence and during science observations that used multiple Proto-NUX filter configurations. For these changes, the telescope was moved to an accessible position on the observing platform, the slider was exchanged by hand, and the observing sequence was then resumed. Communication between the operator at the control computer and the person on the platform was  maintained using handheld radios.

During the first night, the mount alignment procedure was performed and a pointing model was established. We also measured the focus position for each Proto-NUX filter configuration. This was done by pointing the telescope at a bright star and using the tri-Bahtinov mask. The mask provided a sensitive focus diagnostic and also allowed qualitative checks of residual collimation errors. The resulting focus offsets between filters were used as initial values for subsequent nights, reducing the overhead associated with filter changes.

The observing scripts were updated during the campaign to incorporate operational experience from earlier nights. In particular, the sequence of targets, filters, and exposure times was adjusted to reduce overhead while preserving the main goals of the observing campaign (see Section \ref{sec:intro}).

\section{Objectives and observing strategy} \label{sec:objectives}

The Pic du Midi campaign was designed to test Proto-NUX as a complete on-sky system, and to determine which instrumental and atmospheric effects are most important for future Proto-NUX campaigns, for the full NUX facility, and more generally for ground-based NUV observing. For this reason, the observing program included several complementary types of measurements. The main commissioning observations focused on the NUV sensitivity of the system, the atmospheric extinction in the different Proto-NUX filter configurations, the practical calibration of the NUV photometry, and instrumental effects that could affect the photometric accuracy. Sky-background measurements were also obtained during the campaign as supporting diagnostics for the sensitivity estimates and to provide an initial assessment of the NUV observing conditions at Pic du Midi. These observations were complemented by tests on bright sources, which were used to assess the optical performance and to search for ghosting and other internal reflections. A small number of science-demonstration targets were also observed to explore what Proto-NUX can achieve in practice for deep imaging and time-series photometry. In this section we describe the observing strategy and target selection used for these goals. The technical and optical performance derived from these observations is discussed in Section \ref{sec:detector}.

\subsection{Atmospheric extinction and sensitivity}

Atmospheric extinction was measured by observing the flux of one or more stars over a range of airmasses. In principle, an extinction coefficient can be derived from observations at only two airmasses, provided that the atmosphere remains stable. In practice, however, repeated observations over a longer time span give a more reliable measurement. They also make it possible to identify changes in the extinction caused by variable observing conditions, such as clouds, haze, aerosols, or changes in the ozone column. This is especially important for Proto-NUX because the atmospheric transmission across the 300–350 nm band is expected to be strongly wavelength dependent. Different extinction processes dominate in different parts of the band, and their relative contributions can vary with airmass and time. For this reason, extinction measurements were obtained separately in the full-NUX, short-NUX, and long-NUX filter configurations.

A suitable target for this purpose is an open cluster with many unsaturated, high-signal stars in a single Proto-NUX field. This allows the extinction measurement to be based on many sources rather than on a single star. In addition, a wide range of stellar magnitudes is useful, because it provides calibration sources at different signal levels, while saturated stars, low-signal detections, outliers, and variable sources can be excluded. Open clusters can also contain relatively blue stars with strong flux in the 300--350 nm range, increasing the number of usable sources for the NUV extinction measurements.

\begin{figure}
    \centering
    \includegraphics[height=8cm]{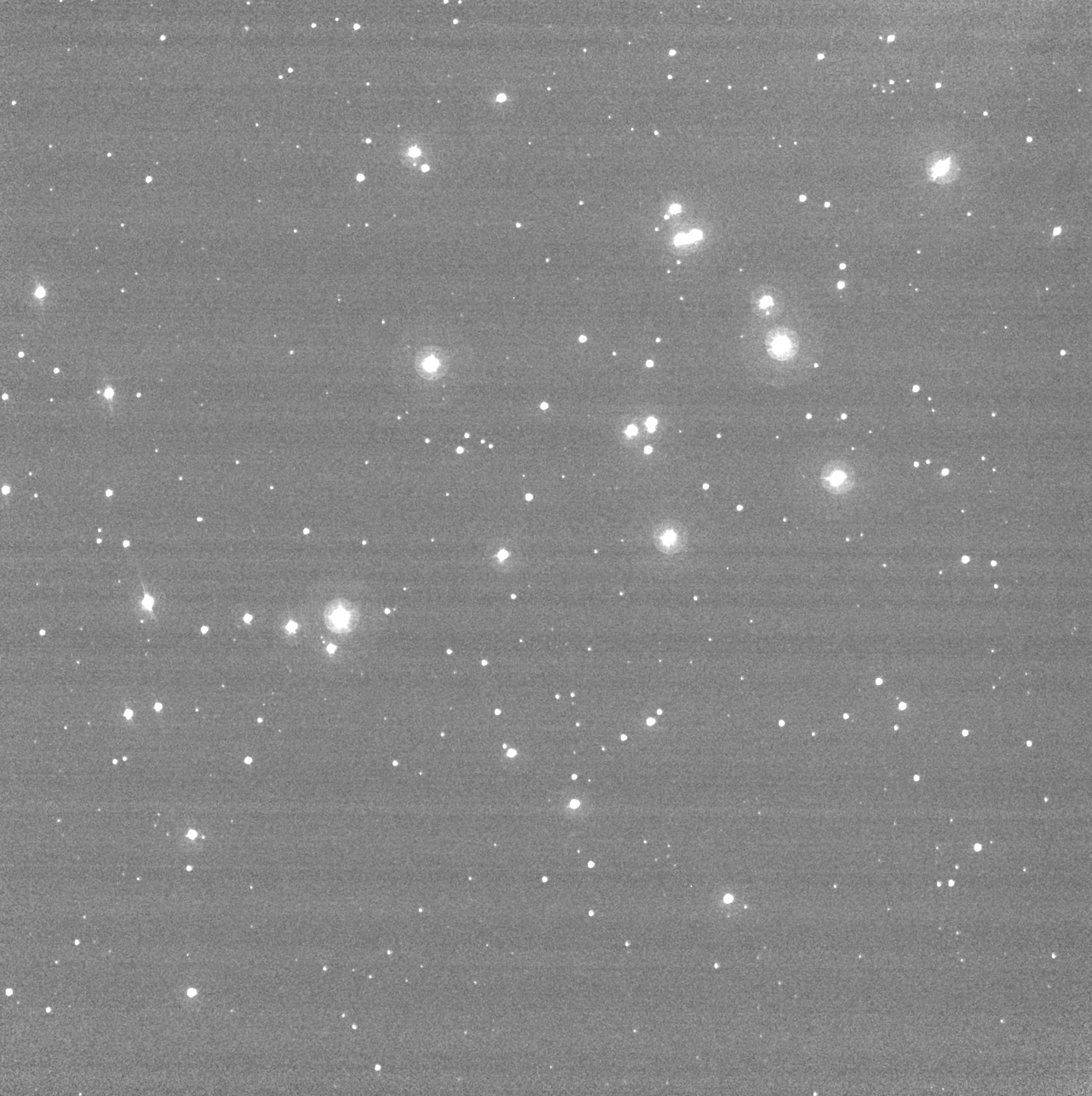}
    \hspace{0.5cm}
    \includegraphics[height=8cm]{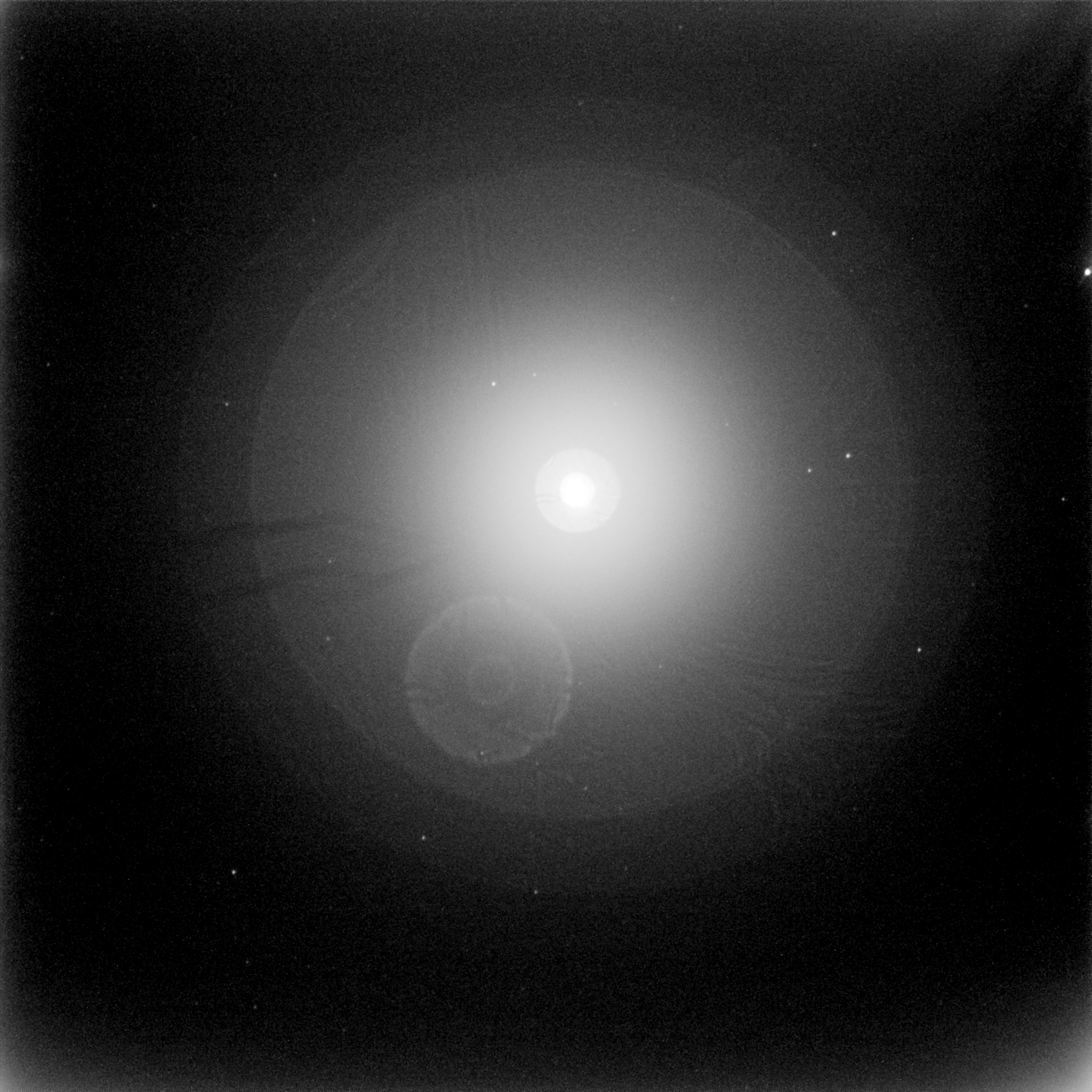}
    \vspace{0.25cm}
    \caption{{\it Left:} Full-NUX image (taken March 17th) of the open cluster M44 obtained during the Pic du Midi campaign. The image shows the on-sky performance of Proto-NUX in the 300--350 nm band, with stellar sources detected across the field. {\it Right:} Full-NUX image of Sirius taken on 17 March, showing the ghosting produced by internal reflections in the optical train. The effect is most visible for very bright stars. }
    \label{fig:m44}
\end{figure}

During the first night of the campaign, we observed NGC~2281
(RA = $06^{\rm h}48^{\rm m}21.8^{\rm s}$,
Dec = $+41^\circ03^\prime36^{\prime\prime}$,
$V\simeq5.4$), which was well placed during the available observing window. From the second night onward, we focused on Messier~44 (RA = $08^{\rm h}40^{\rm m}13.0^{\rm s}$, Dec = $+19^\circ37^\prime16^{\prime\prime}$, $V\simeq3.7$). This cluster provided a longer observing window and could be followed over a wider range of airmasses, making it better suited for the extinction measurements.  M44 also contains known variable stars, such as $\delta$ Scuti pulsators and eclipsing binaries. These sources were not used for the extinction calibration, but they provided possible ancillary targets for NUV time-domain studies. For this cluster, we therefore obtained shorter exposures in addition to the standard 150 s exposures. These shorter integrations reduced saturation of the brightest stars in the field and also allowed us to test whether the known variable stars could be studied in the NUV. The results of these variability studies will be presented in a separate paper. An image taken of M44 in the full-NUX band is shown in Figure \ref{fig:m44}, left.

To determine the sensitivity of Proto-NUX, we observed the same cluster near zenith in each of the Proto-NUX filter configurations. The sensitivity was measured separately for each configuration, because the limiting magnitude depends on the combined effects of atmospheric transmission, instrumental throughput, sky background, detector noise, image quality, and exposure time. These observations were repeated over several nights to measure how the sensitivity changed with observing conditions, in particular transparency and delivered image quality. 

\subsection{Ghosting and red-leak diagnostics}

Reflections between optical surfaces can produce low-level ghost images whose intensity scales with the flux of the source. These ghosts are therefore easiest to detect around bright stars. Measuring their position, shape, and relative brightness gives a direct check of the stray-light behaviour of the optical system. It can also provide constraints on possible throughput losses associated with the absence of anti-reflection coatings on the lenses. Bright-star observations can also used to check for out-of-band transmission in the filter configurations. Such red leak can produce unexpectedly high count rates, especially for red stars, and may in some cases introduce additional ghost features. To investigate ghosting, we observed Sirius (see Fig.~\ref{fig:m44}, right), the brightest stellar point source accessible during the campaign. At the time of our observations, Sirius was at an altitude of $\sim$30 degrees. The star was placed sequentially at the centre of the field of view and near each corner of the detector. This sequence was then repeated for each filter configuration.

\subsection{NUV imaging}

\begin{figure}[htbp]
    \centering
    \begin{subfigure}[b]{0.45\textwidth}
        \centering
        \includegraphics[width=\textwidth]{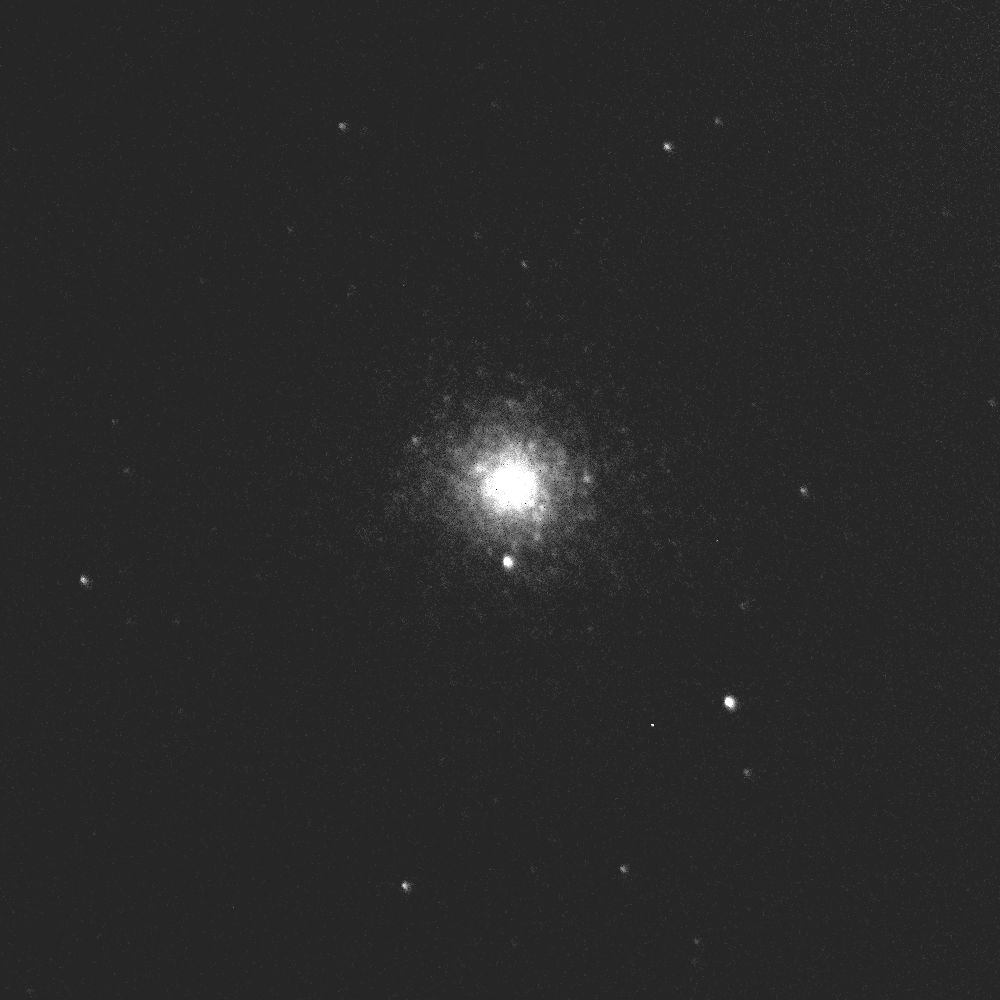}
        \caption{Messier 13, 150 s exposure}
        \label{fig:M13}
    \end{subfigure}
    \hfill
    \begin{subfigure}[b]{0.45\textwidth}
        \centering
        \includegraphics[width=\textwidth]{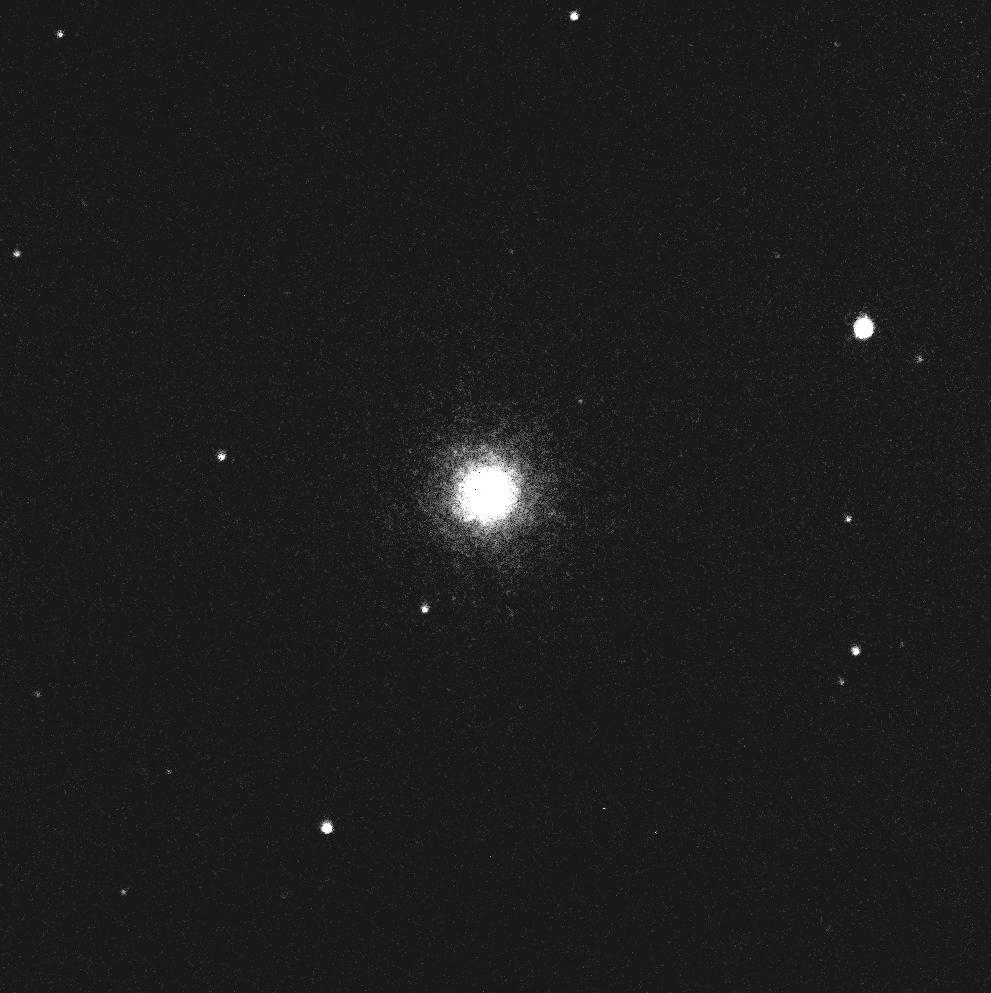}
        \caption{Messier 3, 150 s exposure}
        \label{fig:M3}
    \end{subfigure}
    
    \vspace{1em} 
    
    \begin{subfigure}[b]{0.45\textwidth}
        \centering
        \includegraphics[width=\textwidth]{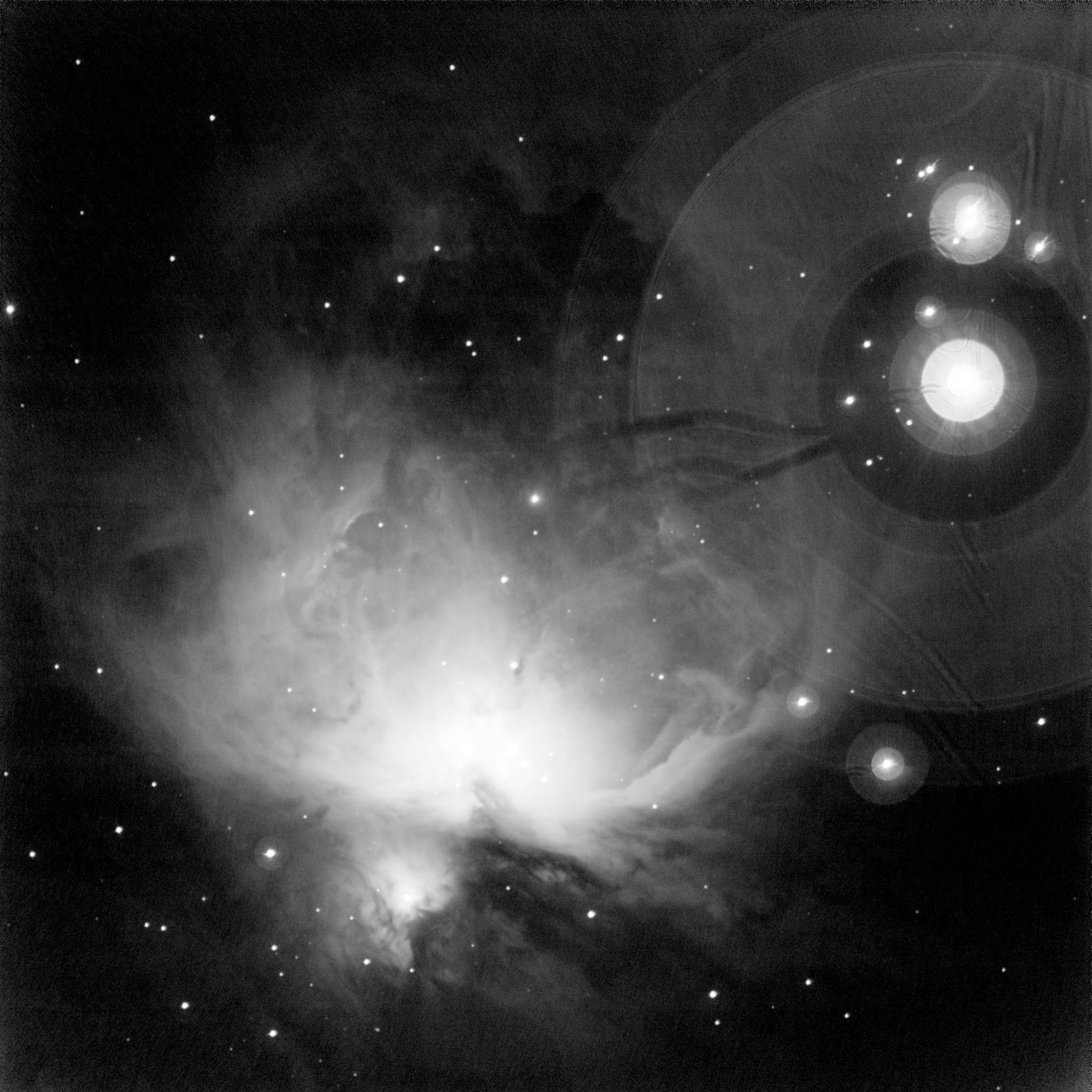}
        \caption{Messier 42, $30 \times 150$  s exposure}
        \label{fig:M42}
    \end{subfigure}
    \hfill
    \begin{subfigure}[b]{0.45\textwidth}
        \centering
        \includegraphics[width=\textwidth]{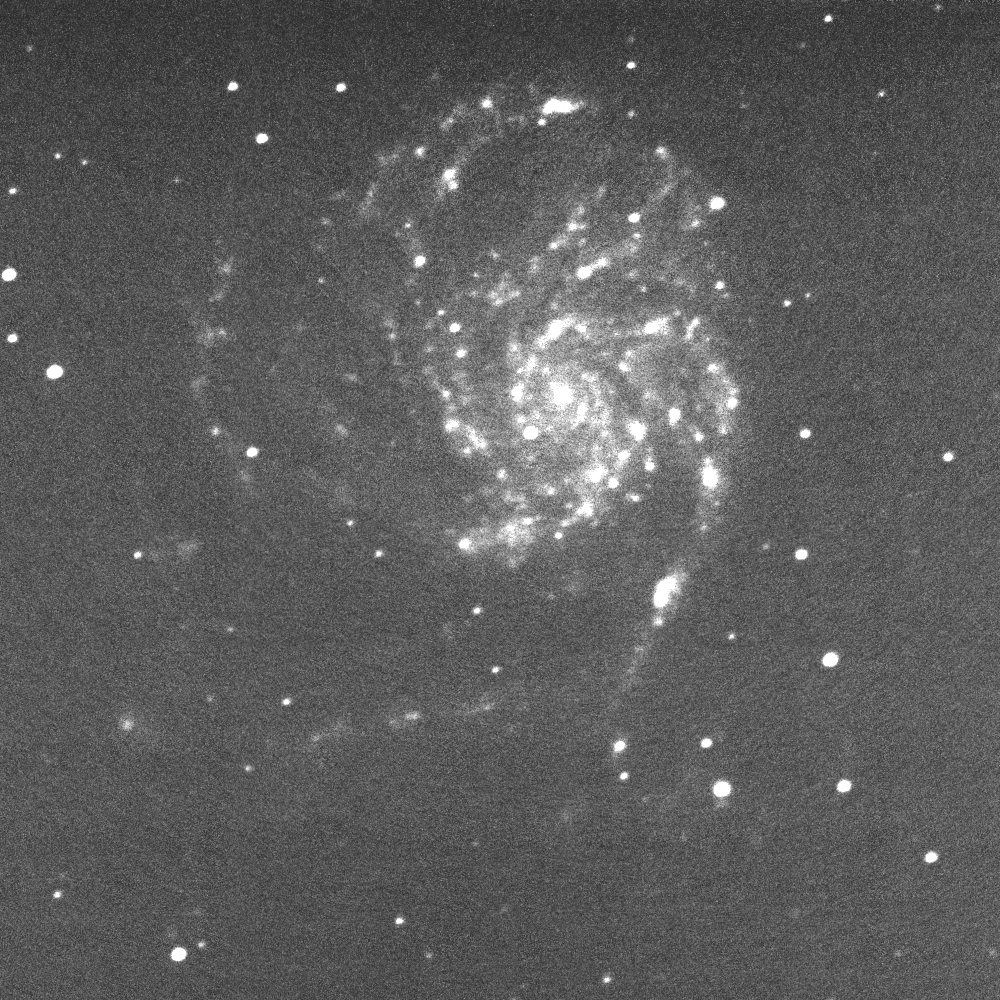}
        \caption{Messier 101, $16 \times 150$ s exposure}
        \label{fig:M101}
    \end{subfigure}
    \vspace{0.25cm}
    \caption{Example images obtained with Proto-NUX during the Pic du Midi campaign. Panels (a) and (b) were taken with the short-NUX filter, while panels (c) and (d) were taken with the full-NUX filter. Panels (a), (b), and (d) are cropped to $27' \times 27'$, whereas panel (c) shows the full Proto-NUX frame of $33' \times 33'$.}
    \label{fig:vier_figuren}
\end{figure}

The scientific potential of ground-based NUV imaging is still largely unexplored (see [\citenum{2024SPIE13094E..30W}]). To test what Proto-NUX can achieve in longer NUV exposures, we observed several extended sources with different astrophysical characteristics, including globular clusters, nebulae, and galaxies (see Fig. \ref{fig:vier_figuren}). These targets were not intended as a complete science sample, but as demonstration observations spanning a range of source types, surface brightnesses, and angular scales. Initial observations were obtained of Messier 42, a bright and extended star-forming region, and of the globular clusters Messier 3 and Messier 13. These targets were chosen to test the performance of Proto-NUX on both extended emission and crowded stellar fields.

During the campaign, the delivered image quality was poorer than expected (see Tab. \ref{tab:seeing}). This reduced the scientific value of observations of dense stellar systems, where source confusion becomes important once the point-spread function is too broad. Atmospheric seeing contributed to the measured image quality, but the comparison between the Proto-NUX and optical images obtained with COX suggests that instrumental effects also played a role in broadening the NUV point-spread function (see Section \ref{sec:imagequality} for details). For this reason, the remaining deep-imaging time was concentrated on a single target, in order to maximize the final depth.

Messier 101, the Pinwheel Galaxy, was selected as the primary deep-imaging target because previous UV observations have shown prominent emission from young stellar populations and star-forming regions in its disk (see, e.g., [\citenum{1997ApJ...481..169W,2005ApJ...619L..71B}]). Its well-defined spiral structure contains numerous O- and B-type stars and extended UV-bright star-forming regions, making M101 a suitable target for Proto-NUX imaging. At the same time, its large angular size and relatively low surface brightness make it a challenging test case for assessing both the point-source and surface-brightness sensitivity of ground-based NUV observations. By integrating on a single target for an extended period, the campaign aimed to measure the achievable depth and to test whether diffuse extragalactic NUV structures can be detected from the ground.

\subsection{Photometric calibration}

Photometric calibration of the Proto-NUX data was based on field stars with known SDSS $u$- and $g$-band magnitudes. For each calibration star, the cataloged $u - g$ color index was compared to a grid of Kurucz stellar-atmosphere models\cite{2003IAUS..210P.A20C}. Synthetic Proto-NUX magnitudes were calculated by convolving the model spectra with the measured system-throughput curves of the instrument. This gives a consistent transformation from SDSS photometry to the Proto-NUX AB magnitude system. A more detailed description of the calibration procedure will be presented in [\citenum{protonuxaapaper}]. The method is most reliable for stars with effective temperatures above approximately 11,000 K. In this regime, the $u - g$ colour remains a monotonic function of temperature, so a given colour corresponds to a unique model temperature. At lower temperatures, the $u - g$ colour–temperature relation becomes degenerate: the same colour can correspond to more than one effective temperature. This introduces an ambiguity in the model selection and reduces the reliability of the inferred Proto-NUX magnitude.

In addition to the field-star calibration, we observed the spectrophotometric standard star Feige 34 as an independent absolute-calibration check. Feige 34 is a hot subdwarf O-type star (sdO) with $V \approx 11.2$ and $B - V \approx -0.34$, and is therefore well suited for NUV calibration because of its strong UV flux and relatively smooth spectral energy distribution\cite{2017ASPC..509..219L}. Its flux distribution is well characterized by spectrophotometric reference spectra, and the star is commonly used for instrument calibration in the optical and UV. The Feige 34 observation therefore provides an independent check on the photometric zeropoint and on possible systematic errors in the field-star calibration.

\subsection{Planetary NUV observations}

Venus and Jupiter were observed several times during the campaign (see Tab.~\ref{tab:Observations}) as bright, extended NUV targets. These observations were included mainly as exploratory tests of how Proto-NUX performs on resolved Solar System objects, rather than as part of a dedicated planetary-science program. Venus is a useful target because its clouds show strong UV contrast \cite{Lee2015,Lee2021}, while Jupiter provides a larger resolved disk and may also offer possibilities for future UV studies of auroral activity \cite{Grodent2015}. Given the delivered image quality during the campaign, the observations were used primarily to assess the practical feasibility of planetary NUV imaging with Proto-NUX, including the photometric stability and delivered image quality on bright, extended sources. They also provided a first check of whether large-scale changes in the NUV appearance of such targets could be detected.

\subsection{NUV exoplanet transit photometry}

Exoplanet transit photometry can show wavelength-dependent transit depths, because atmospheric opacity changes the effective radius of the planet as a function of wavelength\cite{Seager2000,Brown2001}. In the near-infrared, for example, the He I absorption feature at 1083 nm has been used to detect extended and escaping upper atmospheres around close-in exoplanets\cite{Spake2018}..

In the NUV, wavelength-dependent transit signatures are less well explored, but several studies have shown that extended upper atmospheres can produce excess absorption at these wavelengths. This absorption can be caused by high-altitude atomic and ionized species, including metals such as Mg and Fe, and may increase the effective occulting radius of the planet in the NUV\cite{Haswell2012,Cubillos2020}. In addition, UV-driven chemistry can produce small haze particles in the upper atmosphere, which may add scattering or absorption at short wavelengths\cite{Kawashima2018}. These effects make NUV transit observations a challenging but potentially useful probe of extended and escaping exoplanet atmospheres.

To assess the feasibility of NUV transit photometry with Proto-NUX, we observed a transit of WASP-14b on 18 March 2026. WASP-14b is a massive hot Jupiter with a short orbital period of 2.24 days and a relatively large transit depth\cite{joshi2009}, making it a suitable target for testing NUV time-series photometry with Proto-NUX. Its host star, WASP-14, is an F5V star with an apparent magnitude of $V \sim 9.7$. The dataset was intended as a first test of the photometric precision achievable on transit timescales, and to assess whether Proto-NUX could reach the precision needed for future searches for wavelength-dependent transit-depth differences in the NUV.

\begin{sidewaystable}
    \caption{Log of the observations performed at Pic du Midi. In the
third column, FNUX, SNUX, and LNUX refer to the full-NUX, short-NUX,
and long-NUX filter configurations, respectively.}
    \vspace{0.2cm}
    \centering
    \begin{tabular}{|c|c|c|c|c|c|c|l|}
        \hline
        Date &  Object & Filters & Objective & Time & Airmass & Filters & Remarks \\
             &         & Proto-NUX &      &           &        &  COX        & \\
        \hline
        12-3-2026 & NGC 2281 & FNUX & Sensitivity & 21:48-22:46 & 1.04 - 1.1  & - & Very Windy \\
                  &  & SNUX & Extinction & 21:59-4:14 & 1.1 - 5.0 & $u,g,r,i,z$ & Windy until 2:00 \\
                  & M 13 & SNUX & Deep Imaging & 4:28-4:35 & 1.1 & - &  \\
                  & M 3 & SNUX & Deep Imaging & 4:41-4:47 & 1.2 & - &  \\
                  & Moon & FNUX/SNUX/$u$ & Sky Background & 5:06-6:09 & 5.0-10.5 & - &  \\
        \hline
        15-3-2026 & M 31 & FNUX     & Deep Imaging & 20:31-22:15 & 2.4-5.7 & $u,g,r,i$ & \\
                  & M 44 & FNUX     & Sensitivity & 22:49-23:35 & 1.1-1.2 & $L,u,g,r,i$ &Snow on telescope\\
                  &      & LNUX     & Sensitivity & 23:42-00:28 & 1.2-1.3 & & \\
                  &      & $u$        & Sensitivity & 00:33-1:27 & 1.3-1.5 & & \\
                  &      & SNUX     & Sensitivity & 1:33-2:16 & 1.5-1.8 & &\\
                  &      & LNUX/SNUX& Extinction & 2:30-4:58 & 2.0-11.5 & & Alternated every 15 min \\
                  & M 101 & FNUX & Deep Imaging & 5:08-6:33 & 1.1-1.2 & $L$ &  \\
        \hline
        16-3-2026 & Sirius & FNUX/LNUX/SNUX & Ghosting & 19:52-22:21 & 2.0-2.5 & - & \\        
                  & Jupiter & LNUX/SNUX & Egress moon & 21:06-21:25 & 1.1 & $g,r,i,z$ & \\                
                  & M 44 & FNUX     & Sensitivity & 22:42-00:16 & 1.1-1.3 & $u,g,r,i,z$ &\\        
                  &      & FNUX     & Extinction & 00:18-4:53 & 1.3-11 & $u,g,r,i,z$  &\\
                  & M 101 & FNUX     & Deep imaging & 05:02-6:34 & 1.1-1.2 & $g,r,i$  &\\        
        \hline
        17-3-2026 & Sirius & no-filter/LNUX/$u$ & Ghosting & 19:52-20:24 & 2.0-2.1 &  - &  \\
                  & Feige 34 & FNUX/LNUX/SNUX/$u$ & Photometric calibration & 20:51-23:35 & 1.0-1.2 &   $u,g,r,i,z$ & \\
                  & M 42 & LNUX & Deep imaging & 22:02-22:35 & 2.5 &   - & \\
                  & M 45 & LNUX & Deep imaging & 22:45-22:48 & 3.0 &  -  & \\
                  & M 44 & LNUX & Sensitivity & 23:46-1:59 & 1.2-1.7 & $u,g,r,i,z$  & Good seeing \\
                  &      & FNUX/SNUX    & Extinction & 2:30-5:00 & 2.1-16 & $u,g,r,i,z$  & \\
                  & M 101 & FNUX & Deep Imaging & 4:45-6:18  & 1.1 & $L,g,r,i$  & \\
        \hline
        18-3-2026 & Venus & FNUX & UV variability & 19:51-20:10 & 7.4-12.5 & $u,g,r,i,z,y$   & \\
                  & Jupiter & FNUX & UV variability & 20:12-21:43 & 1.1 &   $u,g,r,i,z,y$ & \\
                  & M 101 & FNUX & Deep imaging & 20:47-23:14 & 1.1-1.3 &   $g,r,i$ & \\
                  & WASP-14 & FNUX & Exoplanet Transit & 23:56-6:02 & 1.2-1.6  & $g,r,i$ & Cloudy  \\
        \hline
    \end{tabular}    
    \label{tab:Observations}
\end{sidewaystable}

\section{On-sky performance} \label{sec:detector}

The commissioning data show how Proto-NUX performed under real high-altitude observing conditions. Here we focus on the main issues that shaped the quality of the NUV data: image quality, detector behaviour, and the operational stability of the telescope system. These results also show where the current setup falls short, and what needs to be improved before future observing campaigns.

\subsection{Delivered image quality} \label{sec:imagequality}

The image quality measured in the Proto-NUX bands was worse than expected from the simultaneous optical data alone. We therefore treat the measured stellar widths as the delivered image quality of the full system, rather than as a direct measurement of the atmospheric seeing. The delivered point spread function (PSF) includes a contribution from seeing, but is also affected by optical alignment, focus, chromatic aberration, tracking, detector sampling, and internal reflections.

We evaluated the delivered image quality from point-source images taken with the different Proto-NUX filter configurations, supported by simultaneous COX observations. Although the overlap between the two fields was smaller than intended, the common area still provided a useful reference obtained under the same atmospheric conditions and mount behaviour. The comparison is not a one-to-one benchmark, since the two systems differ in optical design, detector, plate scale, and wavelength response, but it gives valuable context for interpreting the delivered Proto-NUX image quality.

\begin{figure}
    \centering
     \includegraphics[height=7cm]{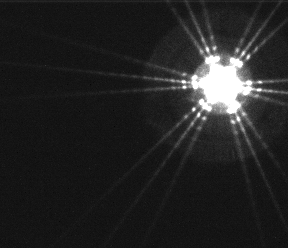}
    \hspace{0.5cm} 
    \includegraphics[height=7cm]{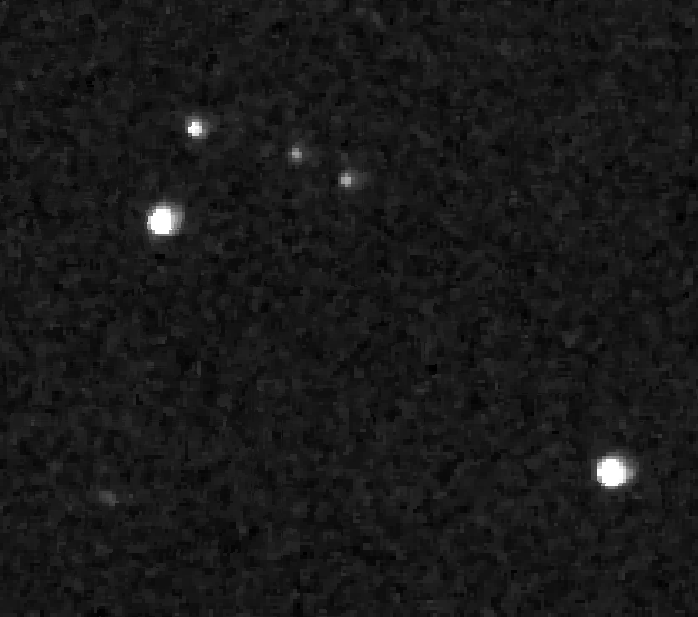}
    \vspace{0.25cm}
    \caption{{\it Left:} Tri-Bahtinov pattern of Sirius obtained with Proto-NUX through the full-NUX filter. The mask was used to set the telescope focus before the observations and to provide a qualitative check of the optical alignment. In a perfectly collimated system, the three diffraction patterns, separated by $120^\circ$, would be identical. The observed differences between them indicate residual collimation errors and chromatic aberration, which contributed to the delivered image quality. {\it Right:} Residual coma observed in Proto-NUX images from 17 March 2026, the night with the best observing conditions of the campaign. The coma-like asymmetry is visible across the field and does not point toward the geometrical centre of the detector, indicating that the optical axis was offset from the detector centre during these observations.}
    \label{fig:coma}
\end{figure}

During the campaign, Proto-NUX was focused on bright stars using a tri-Bahtinov mask (Fig.~\ref{fig:coma}, left). This provided a practical and repeatable focusing procedure, but also showed that the system was not perfectly collimated. The adopted focus should therefore be regarded as the best practical focus for the observing configuration, rather than as a unique optimum valid for all filters. Residual misalignment and chromatic effects across the broad Proto-NUX band can both broaden the PSF, especially toward the shortest wavelengths.

To quantify the image quality under good observing conditions, we used data from 17 March, taken shortly after the telescope had been focused. Stellar FWHM values were measured in the standard COX optical filters ($u$, $g$, $r$, $i$, and $z$) and in the three Proto-NUX filter configurations: short-NUX, long-NUX, and full-NUX. The results are summarised in Table~\ref{tab:seeing}.

\begin{table}[t]
\vspace{0.5cm}
\caption{Measured stellar FWHM values from the best Feige~34 frame obtained on 17 March, after focusing and under good observing conditions. The values include both atmospheric seeing and instrumental broadening. The $u$ band is listed twice because it was observed with both Proto-NUX and with COX. The estimated uncertainty is $\lesssim 0.1^{\prime\prime}$ for all measurements.}
\vspace{0.25cm}
\centering
\begin{tabular}{l c}
\hline
Filter &Measured stellar FWHM \\
\hline
u     & 4.3" \\
g     & 2.3" \\
r     & 1.2" \\
i     & 1.3" \\
z     & 1.4" \\
\hline

u & 5.4" \\
short-NUX  & 4.7" \\
full-NUX  & 4.3" \\
long-NUX  & 4.3" \\
\hline
\end{tabular}
\label{tab:seeing}
\end{table}

The optical data obtained with COX show the expected behaviour: the images in the $r,i,z$ have smaller FWHM than in the $u,g$ bands. COX is a commercially available telescope which are typically not optimized for shorter wavelengths. However, the Proto-NUX images are broader than a simple extrapolation of this trend would suggest. This indicates that the Proto-NUX image quality was not determined by the atmosphere alone. The comparison with COX in the $u$ band is particularly useful here: COX produced sharper images than Proto-NUX by roughly one arcsecond, and in some cases by more. Some of this difference is expected, since seeing becomes worse at shorter wavelengths. However, the observed difference is too large to attribute to seeing alone, and points to an additional contribution from the Proto-NUX system itself.

Within the area sampled by the QHY2020 detector, the stellar profiles are broadly uniform. We do not see clear evidence for strong field curvature, astigmatism, or other large-scale aberrations across the detector. This is encouraging, because it suggests that the present optical train can produce a reasonably flat image over the field tested here. The main limitation was therefore not the spatial uniformity of the PSF, but its width.

The same 17 March data also showed residual coma in the Proto-NUX images (Fig.~\ref{fig:coma}, right). The effect was easiest to identify in these data because they were obtained during the best observing conditions of the campaign, when the atmospheric contribution to the PSF was relatively small. The coma was visible across the full detector field, rather than being confined to one side or corner. In addition, the coma did not appear to point toward the geometrical centre of the detector. This indicates that the optical axis was offset from the detector centre during the observations.

Taken together, the broad Proto-NUX PSF, the comparison with COX, and the residual coma all point to an instrumental contribution to the delivered image quality. The most likely causes are residual collimation errors, a small offset between the optical axis and the detector centre, small focus offsets, and chromatic effects in the NUV optical train. The greater sensitivity of short-wavelength observations to small wavefront errors may also have contributed.  Internal reflections may also have played a role. Ghosting and scattered light mostly affect the wings of the PSF, rather than the core, but they can still make stellar profiles broader and make photometry more difficult, especially near bright sources. The Proto-NUX filter configurations used during this campaign consisted of stacked filters, which may have added further reflecting surfaces and therefore contributed to these effects. 

This interpretation is consistent with the purpose of Proto-NUX. The instrument was designed for the full-NUX band, whereas COX was not optimized for NUV imaging. After proper collimation and focusing, Proto-NUX should therefore be capable of delivering substantially better NUV image quality than was achieved during this campaign. The broad PSF measured during this campaign should therefore be seen as a limitation of the current setup, not as the final performance of the optical design.

\subsection{Detector behaviour}

Detector behaviour was characterised using series of bias and dark exposures. A master bias frame was constructed from the median of multiple bias frames. A master 150 s dark frame was then created by median-combining individual dark exposures, after which the master bias was subtracted to isolate the thermally generated dark signal. As described in Section~\ref{sec:setup}, the detector was operated in dual-readout mode, producing separate low-gain (LGC) and high-gain (HGC) channels. We therefore analysed the two readout regions separately by splitting each frame into its two channel halves.

To convert the measured signal from ADU to electrons, we used gain values of 4.36~e$^{-}$/ADU for the LGC, supplied by the manufacturer, and 1.74~e$^{-}$/ADU for the HGC, estimated from the manufacturer’s gain curve \footnote{See https://www.qhyccd.com/qhy2020-scientific-camera-gsense2020/} (at gain setting 5). The dark current was then calculated from the median value of the bias-subtracted master dark frame as

\begin{equation}
I_{\mathrm{dark}} = \frac{\mathrm{median}(D - B)\cdot g}{t_{\mathrm{exp}}}
\end{equation}

where $I_{\mathrm{dark}}$ is the dark current in e$^{-}$~pixel$^{-1}$~s$^{-1}$, $D$ is the master dark frame in ADU, $B$ is the master bias frame in ADU, $g$ is the detector gain in e$^{-}$/ADU (corrected for upscaling from 12- to 16-bits), and $t_{\mathrm{exp}} = 150~\mathrm{s}$. This gives dark-current values of 0.52 and 0.50~e$^{-}$~pixel$^{-1}$~s$^{-1}$ for the LGC and HGC, respectively.

The similar dark-current values obtained for the two gain channels show that the result is not driven by one of the readout channels. Since we use the median pixel value, the estimate is also not strongly affected by hot pixels. It should therefore be seen as the typical dark current of the detector during the campaign, rather than as a measure of the worst pixels.

A spatial variation in the dark current is visible across the detector. The full-frame median gives a value of about 0.50~e$^{-}$~pixel$^{-1}$~s$^{-1}$, whereas the central 100$\times$100 pixel region gives a lower value of about 0.32~e$^{-}$~pixel$^{-1}$~s$^{-1}$. The higher values near the detector edges may be caused by thermal gradients across the sensor or by increased leakage current close to the readout electronics. The central region is therefore likely a better measure of the detector performance under normal operating conditions.

Overall, the measured dark current is high compared with what is usually expected from deeply cooled scientific detectors. For a typical CMOS sensor, values at this level would normally suggest that the detector was effectively operating closer to about 0--20~$^\circ$C. We did not have an independent temperature measurement for these frames, and we also did not measure the dark current as a function of temperature. This temperature estimate should therefore only be taken as a rough indication.

In the noise budget, the dark current adds a term that scales as $\sqrt{I_{\mathrm{dark}} t}$. This extra noise reduces the sensitivity to faint sources. Compared with a detector with low dark current ($\ll 1$~e$^{-}$~pixel$^{-1}$~s$^{-1}$), the measured level would reduce the limiting depth by roughly one magnitude, although the exact loss depends on the aperture size and observing conditions.

\subsection{Tracking and operational stability}

The mount performed well during the campaign. Once the pointing model had been made, Proto-NUX could be slewed between targets and manually moved for filter changes without losing the model. This was important because the Proto-NUX filters were mounted in separate manual sliders. During normal observing sequences, we found no evidence that tracking was the main cause of the broad NUV stellar profiles discussed in Section~\ref{sec:imagequality}.

Some limitations were still present. The polar alignment was not perfect, and no full correction was applied during the campaign. This left a small altitude-dependent offset in the pointing. At high airmass, this leads to field rotation and possible elongation of the stellar profiles, especially when multiple exposures are combined. These effects were not the dominant limitation in the commissioning data, but they are relevant for accurate photometry and for observations over a wide airmass range.

As noted in Section~\ref{sec:setup}, the incomplete overlap between the fields of Proto-NUX and COX limited the number of stars available for direct NUV-optical comparison. This was sufficient for the commissioning analysis, but future campaigns would benefit from an adjustable mount for COX, so that the optical and NUV fields can be aligned.

Overall, the telescope system was stable enough for the commissioning observations. The main improvements needed are better polar alignment, better control of high-airmass field rotation, and a better way to align the COX field with the Proto-NUX field.

\section{Lessons learned and future improvements} \label{sec:future}

The Pic du Midi campaign showed that Proto-NUX can be deployed and operated as a complete ground-based NUV observing system. The telescope, mount, cameras, filter configurations, calibration procedures, and remote-control setup all functioned together under realistic high-altitude observing conditions. At the same time, the campaign made clear where the present system is not yet optimal. The main improvements needed before future campaigns concern the optical alignment, the filter and coating configuration, the level of automation, and the detector performance.

\subsection{Optical improvements}

The delivered image quality was the main technical limitation of the Pic du Midi data. As discussed in Section~\ref{sec:imagequality}, the broad NUV PSF was probably caused by a combination of residual coma from imperfect collimation or decentring, small focus offsets, chromatic effects, and internal reflections. The on-sky focusing tests with the tri-Bahtinov mask showed that the Proto-NUX optical alignment was not yet optimal during the campaign. The mask was useful not only for finding the best practical focus, but also as a qualitative check of residual collimation errors. Together with the inferred offset between the optical axis and the detector centre, the residual coma seen in the best data makes the next alignment step clear: before the next campaign, the optical train should be re-collimated with explicit checks on the centring of the optical axis, camera tilt, and focus.

Proto-NUX provides several mechanical degrees of freedom that can be used to improve the alignment, including adjustment of the Schmidt corrector and tilt adjustment of the camera. Before the next observing campaign, the optical train should therefore be re-collimated under controlled conditions, with explicit checks on the centring of the optical axis, camera tilt, and focus. This should then be verified on sky using the tri-Bahtinov mask, followed by point-source measurements across the field. A more systematic focus and collimation procedure, repeated for each filter configuration, will make it easier to separate atmospheric seeing from residual instrumental broadening.

The bright-star observations also showed ghosting and scattered light in the current system. These effects do not dominate the PSF core, but they do affect the PSF wings and can complicate photometry around bright sources. Future versions of the system, including the full NUX implementation, should therefore use improved anti-reflection coatings where possible. The optical surfaces and filter configurations should also be reviewed to reduce internal reflections, especially where stacked filters add extra reflecting surfaces.

A larger-format detector will also be needed to test the PSF over a wider field, since the QHY2020 samples only part of the field that the Proto-NUX optics are designed to deliver. These changes will make it possible to separate temporary alignment and implementation issues from any intrinsic limitations of the current optical configuration.

\subsection{Operational improvements}

The campaign also showed that several procedures that were acceptable for a short commissioning run would become limiting for a longer or more remote deployment. The clearest example is the manual operation of the Proto-NUX filter sliders. Manual filter changes introduced overheads, required coordination between the control room and the observing platform, and limited the flexibility of the observing sequences. For a future semi-permanent or remote installation, filter changes should either be automated, or the observing strategy should be designed around a smaller number of manually selected configurations.

The parallel optical telescope was valuable for providing simultaneous optical context, but the fixed piggyback mounting left a significant pointing offset between COX and Proto-NUX. Future campaigns should use a mounting interface that allows COX to be aligned with the Proto-NUX field. This would increase the overlap between the two instruments and make simultaneous colour information, calibration checks, and PSF comparisons more useful.

A future deployment, for example at the site in Spain discussed in Section \ref{sec:spain}, should be designed for fully remote operation. This requires reliable remote access to the control computer, automated power control, environmental monitoring, and a robust weather-safety procedure. Calibration should also be made less dependent on twilight conditions by adding flat-field panels, at least for COX. 

A permanent or semi-permanent installation would also benefit from an automated reduction pipeline. Such a pipeline should produce quick-look images, source catalogues, image-quality measurements, and basic photometric diagnostics during the night. This would make it easier to identify focus problems, transparency changes, tracking issues, or detector anomalies while observations are still ongoing.

The detector dark-current issue identified during the Pic du Midi campaign needs to be resolved before future sensitivity measurements can be considered representative of the final system. This requires additional laboratory testing of the camera, including dark-current measurements as a function of detector temperature, readout mode, and cooling stability. If the high dark current cannot be reduced to the expected level, the detector or its operating configuration will need to be reconsidered.

Finally, as shown in Figure \ref{fig:aponight} (right panel), several indicator lights on the equipment produce a low level of local light pollution. Although this emission is predominantly red and is therefore not expected to affect the Proto-NUX observations, the planned hosting side for the telescope (Section \ref{sec:spain}) will accommodate several other telescopes. To prevent stray light from reaching these instruments, the indicator lights should be covered. In addition, proper cable management is required to prevent cables from becoming entangled during remote observations.

\subsection{Future observing campaign in Spain} \label{sec:spain}

The next major step for Proto-NUX is a longer deployment at a dedicated telescope-hosting site in Spain. A location has been reserved at 'Eye of the Sky' Telescope Hosting\footnote{https://www.telescope-hosting.com/}. The current plan is to move Proto-NUX there in January or February 2027. Before the move, we will address the main issues identified during the Pic du Midi campaign and discussed above.  

The present plan is to operate Proto-NUX in Spain for at least 1.5 years. This longer deployment will make it possible to move beyond the short commissioning-style observations presented in this paper. The first phase will focus on further technical and calibration work. We expect to spend several months repeating and extending the atmospheric-extinction measurements, testing the stability of the NUV zeropoints, monitoring sky background, and verifying the improved optical performance under a wider range of observing conditions. This phase will also be used to refine the observing scripts, remote-control procedures, calibration strategy, and data-reduction pipeline.

After this calibration and technical-validation phase, the emphasis will shift toward science validation. Proto-NUX is not intended to discover large numbers of transients by itself, because its field of view is still much smaller than that planned for the full NUX facility. Instead, the main transient strategy will be rapid or repeated follow-up of hot and fast transients discovered by other surveys. These observations will test how well ground-based NUV data can constrain the early temperature evolution and colour behaviour of such sources, and will help define the observing strategy for the full NUX array.

The Spanish campaign will also allow us to continue the exploratory science tests started at Pic du Midi. In particular, we plan to obtain additional NUV exoplanet-transit observations to assess the achievable time-series precision and to search for wavelength-dependent transit behaviour in suitable systems. Bright Solar System targets will also be considered, including planets and possibly comets, where NUV imaging may provide useful diagnostics.

The observing program for Spain is still being developed, and we welcome collaborations. Proto-NUX can be used for coordinated observations with other facilities, for follow-up of transients and variable sources, and for observations of persistent targets where NUV coverage would add useful information. Interested collaborators are encouraged to contact us to discuss possible targets, observing strategies, or joint campaigns.

\section*{Acknowledgments}

The Proto-NUX project is supported by the NOVA Instrumentation Program in the Netherlands and by a University of Amsterdam Stimuleringsbeurs awarded to R. Wijnands. This research made use of NASA’s Astrophysics Data System Bibliographic Services. The authors also used AI-assisted language tools to refine parts of the manuscript text and improve clarity. The scientific content, analysis, interpretation, and conclusions remain entirely the responsibility of the authors.

\bibliography{report} 
\bibliographystyle{spiebib} 

\end{document}